\documentstyle[12pt,epsfig]{article}                                                    

\newlength{\dinwidth}                                                    
\newlength{\dinmargin}                                                    
\def\lapproxeq{\lower .7ex\hbox{$\;\stackrel{\textstyle                                                    
<}{\sim}\;$}}                                                    
\def\gapproxeq{\lower .7ex\hbox{$\;\stackrel{\textstyle                                                    
>}{\sim}\;$}}                                                    
\def\be{\begin{equation}}                                                    
\def\ee{\end{equation}}                                                    
\def\bea{\begin{eqnarray}}                                                    
\def\eea{\end{eqnarray}}

\begin{document}                                                    
\titlepage                                                    
\begin{flushright}                                                    
IPPP/08/77   \\
DCPT/08/154 \\                                                    
\today \\                                                    
\end{flushright}                                                    
                                                    
\vspace*{2cm}                                                    
                                                    
\begin{center}                                                    
{\Large \bf Soft Diffraction at the LHC\thanks{Presented by M.G. Ryskin at the International Workshop "Hadron Structure and QCD" (HSQCD'2008),
June 30 - July 4, 2008,  Gatchina, Russia.
 }}                                                    
                                                    
\vspace*{1cm}                                                    
V.A. Khoze$^{a,b}$, A.D. Martin$^a$ and M.G. Ryskin$^{a,b}$ \\                                                    
                                                   
\vspace*{0.5cm}                                                    
$^a$ Institute for Particle Physics Phenomenology, University of Durham, Durham, DH1 3LE \\                                                   
$^b$ Petersburg Nuclear Physics Institute, Gatchina, St.~Petersburg, 188300, Russia            
\end{center}                                                    
                                                    
\vspace*{2cm}

\begin{abstract}
\noindent We present a triple-Regge analysis of the available
$pp\to p\ +\ X$ high-energy data accounting for absorptive
corrections. We describe a model for high-energy soft
interactions which includes the whole set of multi-Pomeron ($n\to
m$) vertices, and give predictions for the LHC.
\end{abstract}



\section{Motivation}

There are two main reasons for revisiting soft $pp$ high energy interactions at this time.

{\bf A.} ~To predict soft processes at the LHC we need a reliable model. A detailed analysis performed in the late sixties
\cite{GM1,GM2} showed
that there could be different scenarios for the high-energy
behaviour of the interaction amplitude.

In the `weak coupling' regime the total cross section
$\sigma_t(s\to\infty)\to const$, and in order not to violate  unitarity, and
to satisfy
  the inequality
\begin{equation}
\sigma^{SD}=\int\frac{d\sigma^{SD}}{dM^2}dM^2\, <\, \sigma_t,
\label{eq:1}
\end{equation}
  the triple-Pomeron vertex must vanish when $t\to 0$, that is
$g_{3P}\propto t$ . In this case, the large logarithm coming from the
integration
  over the mass of the system produced in diffractive dissociation ($\int
dM^2/M^2\simeq \ln s$)
is compensated by the small value of the mean momentum transferred
through the Pomeron $\langle t \rangle \propto 1/\ln s$.

On the other hand, for `strong coupling', with $\sigma_t\propto (\ln s)^\eta$ with
$0<\eta\leq 2$, the inequality (\ref{eq:1}) is provided by a small value of the rapidity
gap survival factor $S^2$ which decreases with energy.

The present diffractive data are better described within the `strong
coupling' approach, and here we show predictions for the LHC for this scenario. However, the possibility of the `weak coupling' scenario is
not completely excluded yet.
Therefore, it is quite important
to study the different channels of diffractive dissociation
at the LHC in order to reach
a final conclusion and to fix the parameters of the model for
high-energy soft interactions.

{\bf B.} ~ The second reason for studying soft interactions arises because it may not be an easy task to distinguish the production of
a new object at the LHC when it is accompanied by hundreds other
particles emitted in the same event.
For
the detailed study of the new object, $A$, it may be better to select the few,
very clean, events with the Large Rapidity Gaps (LRG) on either
side of the new object \cite{KMRpr}. That is to observe the exclusive process
$pp \to p + A + p$.  In such a Central Exclusive
  Process (CEP) the mass of $A$ can be measured with very good accuracy
($\Delta M_A\sim 1-2$ GeV) by the missing-mass method by detecting the
outgoing forward protons.  Moreover, a specific $J_z=0$ selection
rule \cite{Jz} reduces the background and also greatly simplifies the spin-parity analysis of $A$.
However, the CEP cross section is strongly suppressed by the small
survival factor, $S^2 \ll 1$, of the rapidity gaps. Thus we need a reliable model of soft interactions to evaluate
the corresponding value of $S^2$.

\section{Triple-Regge analysis accounting for absorptive effects}
The total and elastic proton-proton cross sections
  are usually described in terms of an eikonal model, which automatically
satisfies  $s$-channel
elastic unitarity. To account for the possibility of excitation of the initial proton, 
that is for two-particle intermediate states with the
proton replaced by $N^*$, we use the Good-Walker
formalism \cite{GW}. Already at Tevatron energies the absorptive
correction to the elastic amplitude, due to elastic eikonal
rescattering, is not negligible; it is about 20\% in comparison with the
simple one Pomeron exchange. After accounting for low-mass proton
excitations (that is $N^*$'s in the intermediate states) the correction becomes twice
larger (that is, up to 40\%). Next, in order to describe high-mass diffractive
dissociation, $d\sigma^{SD}/dM^2$, we have to include an extra factor
of 2  from the AGK cutting rules \cite{AGK}. Thus, the absorptive effects
in the triple-Regge domain are expected to be quite large. The previous
triple-Regge analyses (see, for example, \cite{FF}) did not allow for
absorptive corrections and the resulting triple-Regge couplings must be
regarded, not as bare vertices, but as effective couplings
embodying the absorptive effects. Since  the inelastic cross section
(and, therefore, the absorptive corrections) expected at the LHC are more
than twice as large as that observed at fixed-target and CERN-ISR
energies,  the old triple-Regge vertices cannot be used to predict the
diffractive cross sections at the LHC.
Thus, it is necessary  to perform a new triple-Regge analysis
that includes the absorptive effects explicitly.

\begin{figure}[!thb]
\begin{center}
\includegraphics[height=15cm]{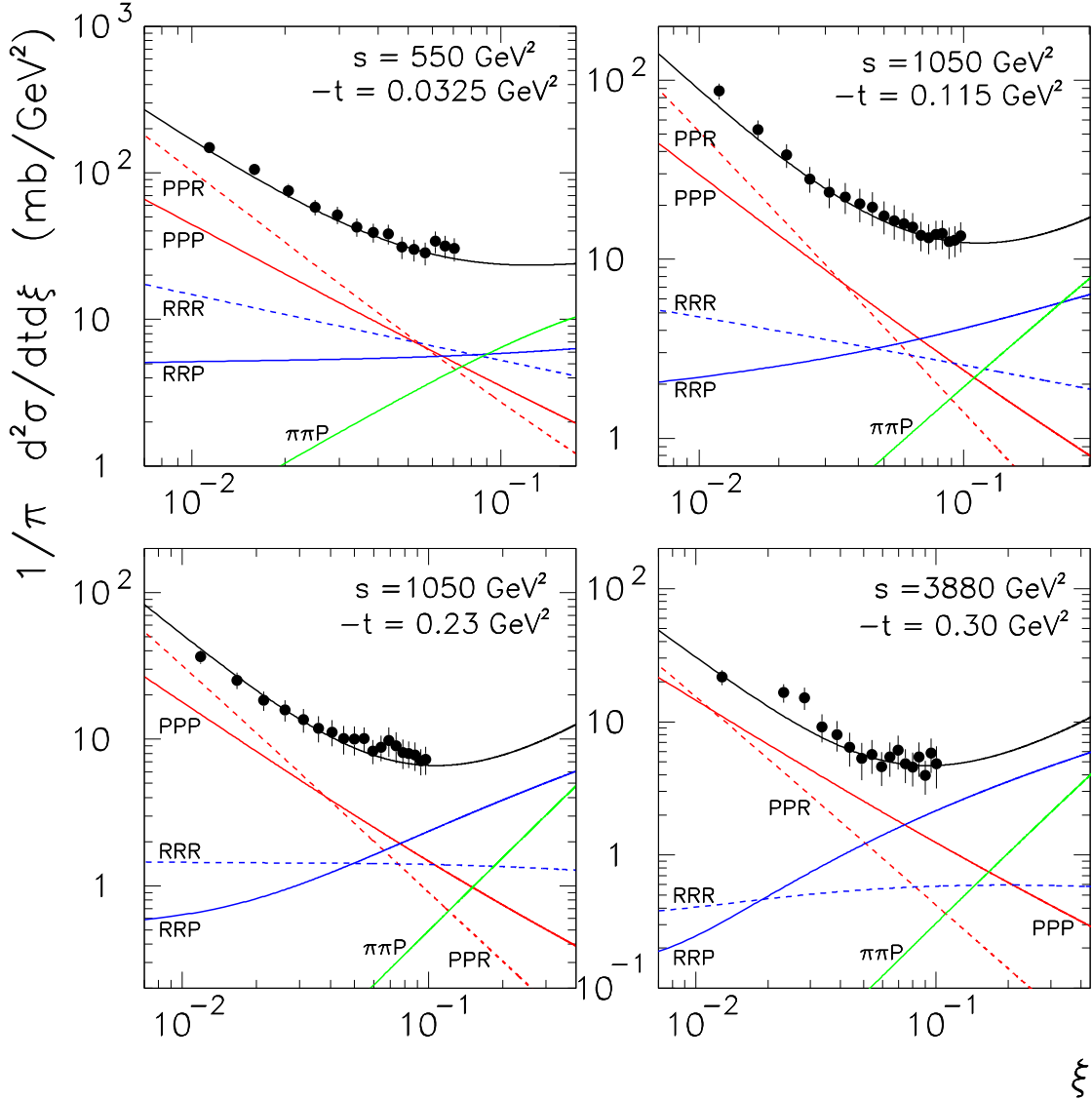}
\caption[*]{The description of the CERN-ISR $pp\to pX$ cross section
$d\sigma/dtd\xi$ data obtained in the strong triple-Pomeron coupling fit, where
$\xi=M^2/s$. The individual triple-Regge contributions are also shown.}
\end{center}
\label{fig1}
\end{figure}

Together with E.G.D.S.Luna, we fitted \cite{LKMR} the fixed-target FNAL, CERN-ISR and Tevatron data that are
available in the triple-Regge region. We
included `$PPP$', `$PPR$', `$RRP$', `$RRR$' and $\pi\pi P$ contributions,
assuming either the `strong' or `weak' coupling scenarios for the
behaviour of the triple-Pomeron vertex. To account for the absorptive
corrections we used a two-channel (Good-Walker) eikonal model, which describes well the
total, $\sigma_t$, and elastic, $d\sigma_{el}/dt$, $pp$ and $\bar
pp$ cross sections.

In the `strong' coupling  case, a good
$\chi^2/$DoF=167/(210-8)=0.83 was obtained. The quality of the description
  to a sample of the data can be seen in Fig.1, where the various
triple-Regge contributions to the cross section are also shown. In
comparison with the old triple-Regge analysis \cite{FF}, we now have about a
twice larger relative contribution of the `$PPR$' term. This is mainly due
to the inclusion in our analysis of the higher-energy
Tevatron data.

Since the absorptive effects are included explicitly,
   the extracted values of the triple-Reggeon vertices are now much closer to the {\it
bare} triple-Regge couplings. In particular, the value
\begin{equation}
g_{PPP}\, \simeq \, 0.2g_N
\label{eq:2}
\end{equation}
is consistent with a reasonable extrapolation of the perturbative BFKL
Pomeron vertex to the low scale region \cite{BRV}; here $g_N$ is the Pomeron-proton coupling.  Note also that these
values of `$PPP$' and `$PPR$' vertices allow us to describe the HERA
data \cite{Jpsi} on inelastic $J/\psi$ photoproduction, $\gamma p\to
J/\psi+Y$, where the screening corrections are rather small.

The `weak' coupling scenario leads to a larger $\chi^2/$DoF=1.4 and to a
worse description of the $\gamma p\to J/\psi+Y$ process at the lowest
values of $t$. At the LHC energy the `weak' coupling fit predicts about 3 times smaller inclusive cross section
$d\sigma^{SD}/dtdM^2$ at $\xi=M^2/s=0.01$ in comparison with that predicted in the `strong'
coupling case.

\section{Model with the whole set of multi-Pomeron vertices}
Since the triple-Pomeron vertex (\ref{eq:2}) turns out to be rather large,
the contribution of the so-called `enhanced' diagrams, with a few
vertices,
  is not negligible. Moreover, we cannot expect that more complicated
multi-Pomeron interactions, driven by the $g^n_m$ vertices, which
describe the transition of $n$ to $m$ Pomerons, will not affect the
final result.  It  looks more reasonable to assume that $g^n_m\propto
\lambda^{n+m}$ than to assume that $g^n_m=0$ for any $n+m>3$. Thus we
need a model which accounts for the possibility of multi-Pomeron interactions
(with arbitrary $n$ and $m$).

While the eikonal formalism describes the rescattering of the incoming
fast particles,  the enhanced multi-Pomeron diagrams
represent the rescattering of the intermediate partons in the ladder
(Feynman diagram) which describes the Pomeron-exchange amplitude.
The ladder-type Pomeron amplitude may be generated by the evolution equation
(in rapidity, $y$, space)
\begin{equation}
\frac{d\Omega(y,b)}{dy}\,=\,
\left(\Delta+\alpha'\frac{d^2}{d^2b}\right)\Omega(y,b)
\label{eq:3}
\end{equation}
where $b$ is the two-dimensional vector in impact parameter
space and $\alpha'$ is the slope of the Pomeron trajectory. $\Delta$ is
the probability to emit new intermediate partons within unit
rapidity interval; the Pomeron intercept
$\alpha(0)=1+\Delta$.
The solution of (\ref{eq:3}),
$\Omega=\Omega_0\exp(\Delta y-b^2/4\alpha' y)/4\pi\alpha' y$,
is the opacity (at point $y,b$), corresponding to the incoming
particle placed at $b=0$ and $y=0$.

It looks natural to account for absorptive effects by including
on the r.h.s. of (\ref{eq:3}) the factor
$\exp(-\lambda\Omega_i/2-\lambda\Omega_k/2)$, where the subscripts $i$ and
$k$ denote the opacities of the beam and the target incoming hadrons. We thus have the evolution equations
\begin{equation}
\frac{d\Omega_i(y,b)}{dy}\,=\, \exp(-\lambda\Omega_i(y,b)/2-\lambda
\Omega_k(y',b)/2)
\left(\Delta+\alpha'\frac{d^2}{d^2b}\right)\Omega_i(y,b)
\label{eq:4}
\end{equation}
\begin{equation}
\frac{d\Omega_k(y',b)}{dy'}\,=\,\exp(-\lambda\Omega_i(y,b)/2-\lambda
\Omega_k(y',b)/2)
\left(\Delta+\alpha'\frac{d^2}{d^2b}\right)\Omega_k(y',b)\; ,
\label{eq:5}
\end{equation}
where here $y'=\ln s -y$.
The
coefficient $\lambda$ accounts for the fact that the absorptive cross
section for the intermediate parton, $c$, may be different from that for the
incoming particle (proton). Since the equations (\ref{eq:3}-\ref{eq:5})
describe the interaction amplitude and not the cross section we have a
factor $1/2$ in the exponent.
In terms of multi-Pomeron vertices, the absorption factors,
$\exp(-\lambda\Omega_i(y,b)/2-\lambda\Omega_k(y',b)/2)$, in
(\ref{eq:4},\ref{eq:5}) correspond to
\begin{equation}
g^n_m\,=\, n\cdot m\cdot\lambda^{n+m-2}g_N/2\;\;\;\;\;\;\;\;\;
\mbox{for $n+m\geq 3$}\, .
\label{eq:g}
\end{equation}
Note that, since the intermediate parton may be absorbed by the
interaction
with the particles (partons) from the wave function of both beam or target
hadron, we now need to solve the two equations, (\ref{eq:4})
and (\ref{eq:5}). This is done iteratively.
The resulting solution solution can then be used in the eikonal formulae
to determine all soft $pp$ interactions.

In comparison with our previous model \cite{KMRs1}, we now include a
non-zero slope ($\alpha'\neq 0$) of the Pomeron trajectory.
The incoming proton wave function is described by three Good-Walker
eigenstates, that is we use a 3-channel eikonal for the rescattering
of fast particles. The transverse size squared of each eigenstate is proportional to the corresponding absorptive
cross section; $R^2_i\propto \sigma_i$.  That is we assume that the parton density
at the origin is the same for each eigenstate. The shape of the
Pomeron-nucleon vertex is  parametrised by the form factor
$V(t)=\exp(dt)/(1-t/d_1)^2$, whose Fourier transform, $V(b)$, plays the
role of the initial conditions for $\Omega(y=0,b)$.

Next, we allow for four different $t$-channel states, which we label $a$: one for the
secondary Reggeon ($R$) trajectory and three Pomeron states ($P_1, P_2, P_3$) to mimic the BFKL
diffusion in the logarithm of parton transverse momentum,
$\ln(k_t)$ \cite{Lip}. To be precise, since the
BFKL Pomeron \cite{bfkl} is not a pole in the complex $j$-plane, but a
branch cut,
we approximate the cut by three  $t$-channel states of a different size.
The typical values of $k_t$ in each of the three states is about $k_{t1}\sim 0.5$ GeV,
$k_{t2}\sim 1.5$ GeV and $k_{t3}\sim 5$ GeV.
Thus the system (\ref{eq:4},\ref{eq:5}) is replaced by
\begin{equation}
\frac{d\Omega^a_i(y,b)}{dy}\,=\, \exp(-\lambda\Omega_i(y,b)/2-\lambda
\Omega_k(y',b)/2)
\left(\Delta^a+\alpha'\frac{d^2}{d^2b}\right)\Omega_i^a(y,b)
+V_{aa'}\Omega^{a'}_i
\label{eq:6}
\end{equation}
\begin{equation}
\frac{d\Omega^a_k(y',b)}{dy'}\,=\,
\exp(-\lambda\Omega_i(y,b)/2-\lambda \Omega_k(y',b)/2)
\left(\Delta^a+\alpha'\frac{d^2}{d^2b}\right)\Omega^a_k(y',b)
+V_{aa'}\Omega^{a'}_k\; .
\label{eq:7}
\end{equation}
The transition factors $V_{aa'}$ were fixed by properties of the BFKL equation.
In the exponents, the opacities $\Omega_i$
($\Omega_k$) are actually the sum of the opacities $\Omega^{a'}_i$
($\Omega^{a'}_k$) with corresponding coefficients (see \cite{KMRs2} for
more details).

Clearly the number of parameters in such a model is large.
Therefore, instead of a straightforward fit of the data, we adjust the
majority of parameters in reasonable intervals and 
demonstrate that such a model allows us to reproduce all the available data on
diffractive cross sections, $\sigma_t,\, d\sigma^{el}/dt,\,
\sigma^{SD}_{\rm low\, mass\, diss^n},\,
d\sigma^{SD}/dM^2$. The quality of the description is demonstrated
in Fig.2, where we also present the prediction for elastic cross section at
the LHC energy $\sqrt{s}=14$ TeV.

\begin{figure}
\begin{center}
\includegraphics[height=18cm]{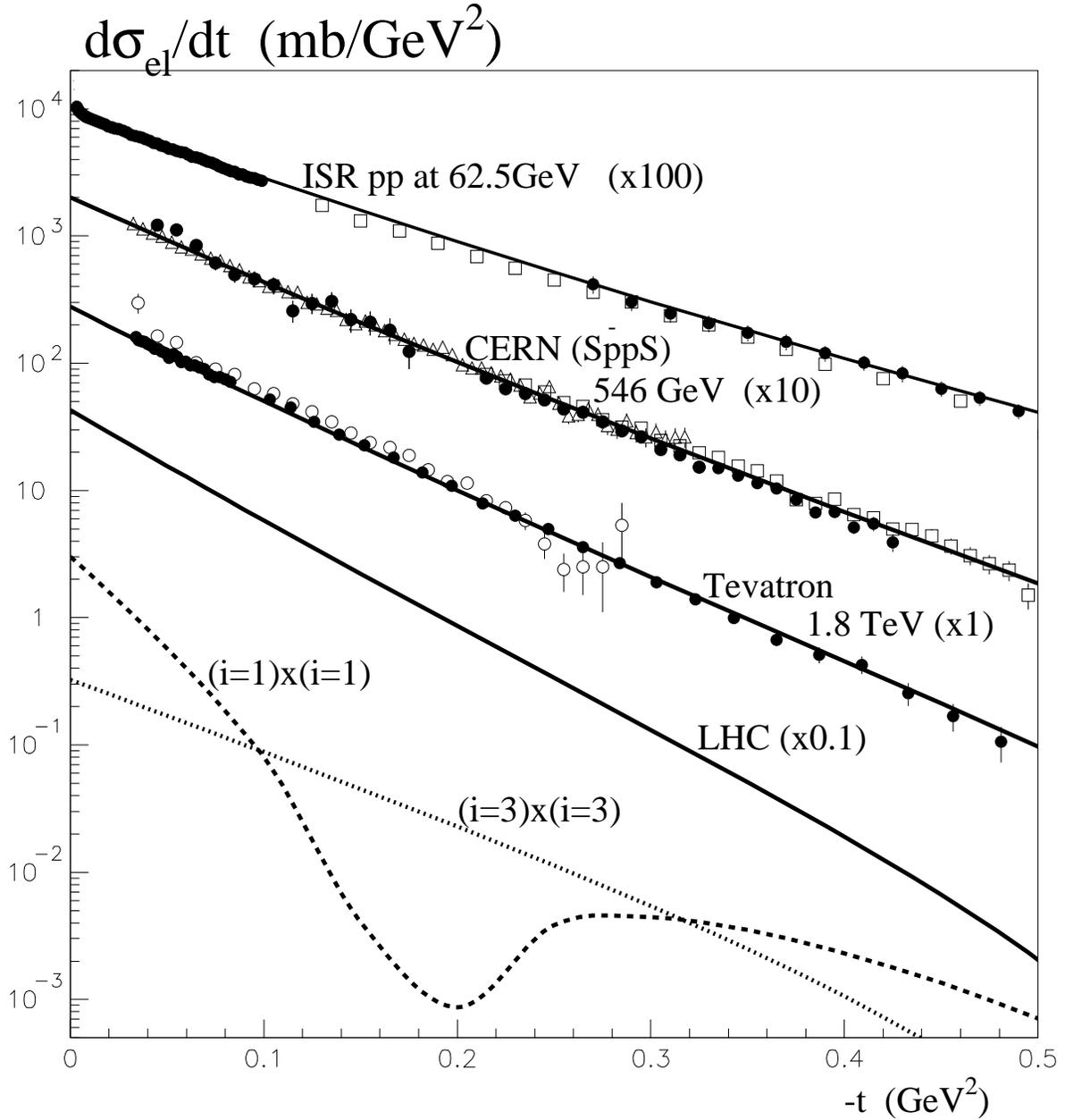}
\caption[*]{The description of the $t$ dependence of the elastic $pp$ cross section data. The
dashed and dotted lines are the contributions at the LHC energy from  the elastic
scattering of the largest size ($i=1$) and the smallest size ($i=3$)
components. Interestingly, we see that the complete set of eikonal contributions gives a smooth prediction for the elastic cross section in this $t$ interval at the LHC energy.} \end{center}
\label{fig2}
\end{figure}

The values of the parameters that we  use are: $d=0.15$ GeV$^{-2}$, $d_1=1.5$ GeV$^2$,
$\lambda=0.25$, $\alpha'_R=0.9$ GeV$^{-2}$, $\alpha_R(0)=0.6$ and, for each of 3 components of the Pomeron, $\Delta^a=0.3$. The Pomeron
intercepts are consistent with the expectations of resummed NLL BFKL which gives
$\omega_0\equiv \alpha_P (0)-1\sim 0.3$ practically independent of the scale
$k_t$ \cite{BFKLnnl}. The slopes of the Pomeron trajectories are:
$\alpha'_{P_1}=0.05$ GeV$^{-2}$ for the large size Pomeron component,
$\alpha'_{P_2}=0.05/9$ GeV$^{-2}$ for the second component
($\alpha'\propto 1/k^2_t$), and
$\alpha'_{P_3}=0$ for the smallest size Pomeron component.

Ths model, which we have tuned to describe the available data, allows us to predict, in principle, all features of soft $pp$ high energy interactions. Let us consider
the cross section for beam particle diffractive dissociation (with a rapidity
  gap up to $y$). First recall the usual eikonal relations 
$$ \sigma_t=2\int d^2b(1-e^{\Omega(b)/2});\;\;\;\;\sigma_{el}=\int d^2b(1-
e^{-\Omega(b)/2})^2;\;\;\;\; \sigma_{inel}=\int
d^2b(1-e^{-\Omega(b)}).$$
For each impact parameter point $b$, the desired cross section for  single
dissociation is proportional (i) to the elastic parton $c-k$ cross section
$(1-\exp(-\lambda\Omega_k(y,b)/2))^2$; (ii) to the probability to find
the intermediate parton $c$ in the interval $dy$, that is
$\Delta\exp(-\lambda\Omega_i/2-\lambda\Omega_k/2)$; (iii) to the
amplitude $\Omega_i$ of the parton $c~-$ beam interaction; (iv) to the
gap survival factor $S^2(b)=\exp(-\Omega(Y,b))$ (where $Y=\ln s$). The
resulting cross section reads $$\frac{d\sigma^{SD}}{dy}\,=\, N\int
d^2b_1 d^2b_2 (1-e^{-\lambda\Omega_k(y,b_1)/2})^2\Delta
e^{-\lambda\Omega_i(Y-y,b_2)/2-\lambda\Omega_k(y,b_1)/2}~~\times $$
\begin{equation}
\times~~\Omega_i(Y-y,b_2)e^{-\Omega(Y,\vec{b}_1+\vec{b}_2)}\, ,
\label{eq:sd}
\end{equation}
where $b_1$ ($b_2$) are the coordinates in
the impact parameter plane with respect to the target (beam) hadron.
Expression (\ref{eq:sd}) must be averaged over the components
(that is the Good-Walker eigenstates) of the beam and target hadrons and summed 
over the different $t$-channel states ($R,P_1,P_2,P_3$) with
appropriate normalisation factors $N$.
The energy behaviour of the diffractive cross sections, and of the
multiplicities of the secondaries produced by the $t$-channel Pomeron
components of different sizes, are shown in Fig.3.

\begin{figure}
\begin{center}
\includegraphics[height=18cm]{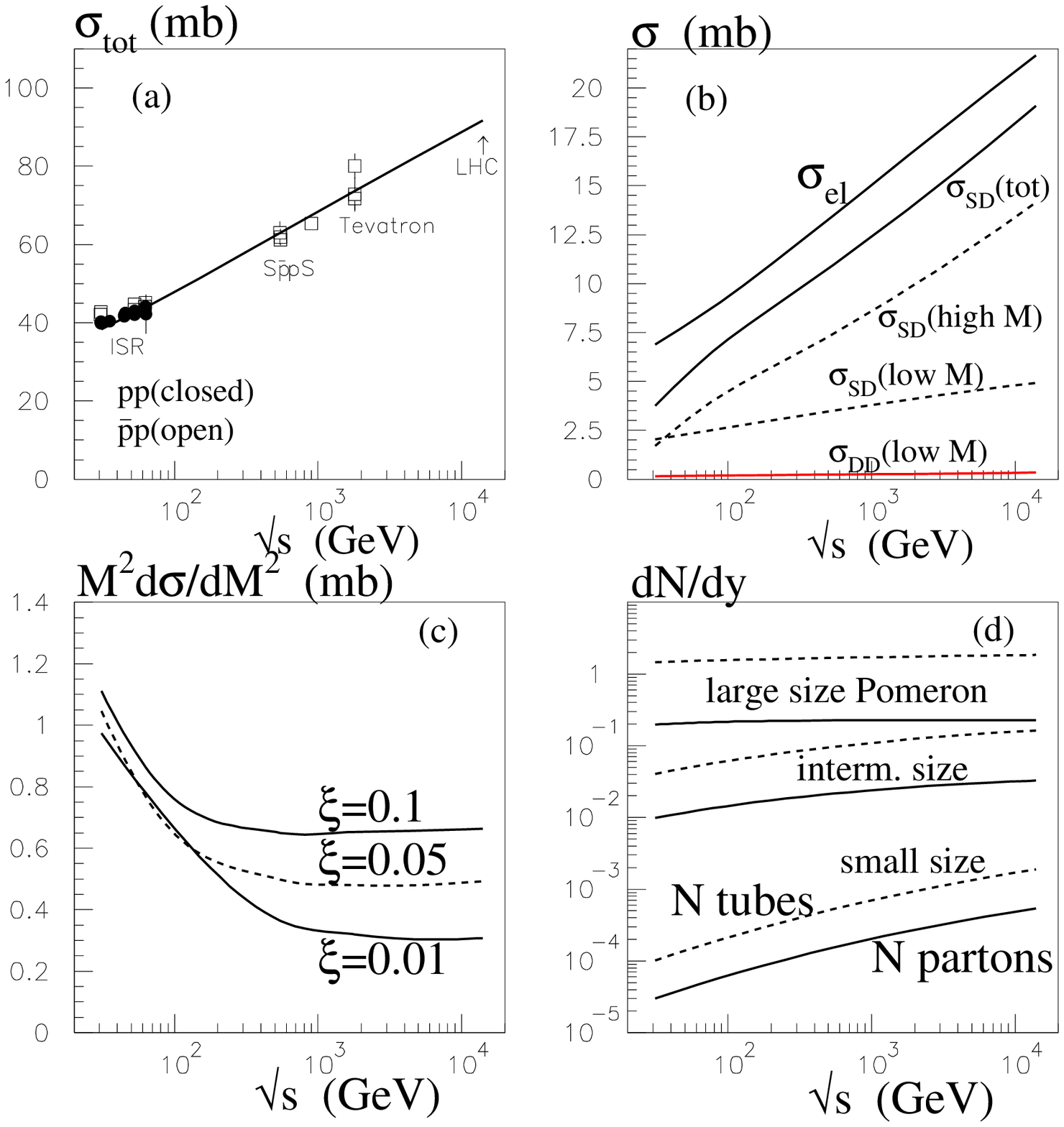}
\caption[*]{The energy dependence of (a) the total, (b) the elastic and
diffractive dissociation, $pp$ cross sections and (c) the cross sections of
dissociation to a fixed $M^2$ state, where $\xi=M^2/s$. Plot (d) shows the parton
multiplicity (solid lines) and the number of `colour  tubes' (dashed)
produced by the Pomeron components of different size.}
\end{center}
\label{fig3}
\end{figure}

Note that, starting with the same `bare' intercepts ($\Delta=0.3$), after the absorptive correction, the contribution of the large-size
component becomes practically
flat, while  the small-size contribution, which is much less affected by
the absorption, continues to grow with energy. Such a behaviour is
consistent with  experiment, where the density of low $k_t$
secondaries is practically saturated, while the probability to produce a hadron
with a large transverse momentum (say, more than 5 GeV) grows with the
initial energy.

\section{Conclusion}
  After accounting for absorptive effects the new
triple-Regge analysis leads to a rather large triple-Pomeron vertex
$g_{3P} = \lambda g_N$ with $\lambda \geq 0.2$. This indicates that in order to describe diffractive processes at
the LHC we must use a model which includes multi-Pomeron interactions.
We construct such a model in which the absorption of the intermediate partons is
described by conventional-like factors $\exp(-\lambda\Omega)$. In terms of
the multi-Pomeron vertices, this corresponds to
  $g^n_m=nm\lambda^{n+m-2}g_N/2$. The model reasonably
well reproduces available diffractive data on $\sigma_t,\,
d\sigma^{el}/dt,\,  d\sigma^{SD}/dM^2$ using the parameters:
$\Delta=0.3$, $\alpha'_{Pom}=0.05$ GeV$^{-2}$, $\lambda=0.25$. It thus leads to the
hope that there is a smooth matching between the perturbative QCD treatment of the
Pomeron and the multi-Pomeron description of soft interactions.

Since the screening corrections caused by the `enhanced' multi-Pomeron
  diagrams, that is by the high-mass dissociation, slow down the growth of
  the cross section with energy,  the model predicts a  relatively low
total
cross section at the LHC energy, namely $\,\sigma_t(LHC)\simeq 90$
mb~\cite{KMRs1,KMRs2}.   Predictions for LHC soft interactions are shown in Figs. 2 and 3.

\end{document}